\documentclass[final,5p,times,twocolumn]{elsarticle}

\usepackage{color,amssymb}
\usepackage[colorlinks, linkcolor=blue, anchorcolor=blue, citecolor=blue]{hyperref}

\usepackage{amsmath} 
\usepackage{multicol} 
\usepackage{graphicx} 
\usepackage{nameref}
\usepackage{cleveref}
\newcommand*{\fullref}[1]{\hyperref[{#1}]{\autoref*{#1} \nameref*{#1}}} 
\journal{Physics of the Dark Universe}

\begin{document}

\begin{frontmatter}


\title{The influence of cosmological constant on light deflection in rotating spacetimes via the generalized Gibbons-Werner method}


\author[first,second]{Yang Huang}
\author[first,second]{Xiangyun Fu*}
\author[first,second]{Zhenyan Lu}
\author[first,second]{Xin Qin}

\affiliation[first]{organization={School of Physics and Electronic Science},
            addressline={Hunan University of Science and Technology}, 
            city={Xiangtan},
            postcode={411021}, 
            state={Hunan},
            country={China}}
\affiliation[second]{organization={Key Laboratory of Intelligent Sensors and Advanced Sensing Materials of Hunan Province},
            addressline={Hunan University of Science and Technology}, 
            city={Xiangtan},
            postcode={411021}, 
            state={Hunan},
            country={China}}

            
\begin{abstract}
    Recently, we proposed a generalized Gibbons-Werner (GW) method for analyzing particle trajectories in rotating spacetimes, regardless of their asymptotic behavior [Huang \textit{et al.}, \href{https://iopscience.iop.org/article/10.1088/1475-7516/2024/01/013}{J. Cosmol. Astropart. Phys. 01(2024), 013}]. Using this method, we examine the impact of the cosmological constant ($\Lambda$) on the light deflection in rotating spacetimes within the framework of Kerr-de Sitter (KdS) geometry. Although Sultana previously calculated the deflection angle of light in KdS spacetime, our study advances this research in three aspects: (i) Orbit solution---the light trajectory is derived by directly solving the original equation of motion (EOM) without intermediate processes. (ii) Positions of the source and observer---the finite distances of the source and observer from the lens are explicitly considered, avoiding approximations. (iii) Staticity of the source and observer---the Randers optical space is employed to resolve the staticity constraint. Through these refined considerations, we obtain a novel expression for the deflection angle of light in KdS spacetime, accurate to second-order in $\Lambda$, as well as in the mass (M) and spin parameter (a) of the central body. Furthermore, we discuss the discrepancies between our results and previous expressions. Finally, we evaluate the observational implications of our corrections relative to Sultana's work in the lensing systems of the Sun and Sgr A*, and show that they may become observable with forthcoming high-precision astronomical measurements.
\end{abstract}

\begin{keyword}
deflection angle \sep cosmological constant \sep Kerr-de Sitter spacetime

\end{keyword}

\end{frontmatter}



\section{Introduction}\label{introduction}

The accelerated expansion of the Universe has been substantiated by the observation of Type Ia supernovae \cite{caldwell1998cosmological,perlmutter1999measurements,astier2005supernova}, large-scale cosmic structures \cite{abazajian2004second,abazajian2005third}, and anisotropies in the cosmic microwave background \cite{spergel2003first}. To account for this phenomenon, various dark energy models have been proposed by modifying the matter term of Einstein’s field equations \cite{sahni2000case,zlatev1999quintessence,caldwell2002phantom,feng2005dark,cai2016dark,mazumdar2001assisted,deffayet2002accelerated,bento2004revival}, with the simplest among them being the cosmological constant model \cite{sahni2000case}.

The gravitational deflection of light is a powerful tool in astrophysics and cosmology, inspiring extensive research into its applications for probing dark energy \cite{rindler2007contribution,sereno2008influence,sereno2009role,bhadra2010gravitational,arakida2012effect,sultana2013contribution,ishihara2016gravitational,arakida2018light,li2020thefinitedistance,takizawa2020gravitational,takizawa2022gravitational} as well as various other aspects \cite{ditta2023testing,ditta2023particle,ditta2023particle2,ditta2023particle3,mushtaq2024weak,ditta2025plasma,mushtaq2025analysis,ditta2025studying}. A troublesome aspect in investigating the influence of $\Lambda$ on light deflection arises from the fact that the de Sitter spacetime is asymptotically non-flat.

The deflection angle of light in KdS spacetime was first given by Sultana in 2013 \cite{sultana2013contribution} based on the scheme proposed by Rindler and Ishak \cite{rindler2007contribution}, with the result
\begin{equation}
    \begin{aligned}
        \delta = & M\cdot \frac{4}{b} +\Lambda \cdot \frac{5\pi b^{2}}{32} +M^{2} \cdot \frac{15\pi }{4b^{2}} -Ma\cdot \frac{4}{b^{2}} -M\Lambda \cdot b\left(\frac{75\pi ^{2}}{512}\right. \\
         & \left. +\frac{101}{96}\right) +a\Lambda \cdot \frac{5\pi b}{16} -\frac{a\Lambda }{M} \cdot \frac{b^{2}}{6} -\frac{\Lambda }{M} \cdot \frac{b^{3}}{6} +\mathcal{O}\left( \Lambda ^{2}\right) ,
        \end{aligned}
    \label{DAKdSSultana}
\end{equation}
which keeps second order terms in $M/b$, $a/b$, $\Lambda b^2$, and $\phi_0$. Here $b$ is the impact parameter, and $\phi_0\ll 1$ (sufficiently small) is the approximate azimuthal coordinate of the source or observer.
   
Recently, the GW method, a powerful approach for calculating the deflection angle of particles from the geometric perspective, has attracted considerable attention \cite{jusufi2016light,ishihara2017finite,jusufi2017light,jusufi2017rotating,jusufi2017effect,sakalli2017hawking,jusufi2017lightdeflection,goulart2017phantom,jusufi2018effect,jusufi2018deflection,ono2018deflection,jusufi2018gravitational97,jusufi2018conical,jusufi2018deflectionrotating,ovgun2018light,ovgun2018shadow,jusufi2019gravitational,ovgun2019exact,ovgun2019weak,javed2019deflection,javed2019effect100,de2019weak,kumar2019shadow,javed2019effect,javed2020weak,javed2020weak135,javed2022weak,javed2022effect,mustafa2022shadows,gao2024microlensing,qiao2023gravitational,jusufi2019distinguishing,crisnejo2019gravitational,crisnejo2019higher,li2020gravitational,li2020finite,li2021kerr,li2021deflection,huang2022generalized,huang2023extending,huang2023finite,pantig2024imprints,sucu2024effect,fu2021weak,zhang2021geometrization,zhang2025global,gao2023gravitational,liu2024gravitational,guo2022bounce,al2024regular,mo2025testing}. This method was initially proposed by Gibbons and Werner in 2008 \cite{gibbons2008applications}, and has since been continuously developed by researchers. The original version of GW method is only applicable to the light in static spherically symmetric and asymptotically flat spacetimes. It has now been extended to the massless and massive particles in various types of spacetimes \cite{ishihara2016gravitational,li2020thefinitedistance,werner2012gravitational,ono2017gravitomagnetic,crisnejo2018weak,jusufi2018gravitational,huang2024generalized}. In our previous paper, we proposed a generalized GW method for rotating spacetimes, even if they are asymptotically non-flat \cite{huang2024generalized}. Owing to the applicability of our generalized GW method to de Sitter spacetime, we adopt it in this paper to study the influence of $\Lambda$ on light deflection in rotating spacetimes, specifically within the context of KdS spacetime. 

This paper is structured as follows: In Sec.~\ref{3problems}, we present our schemes from three aspects based on the generalized GW method. In Sec.~\ref{CalDefAngle}, we calculate the deflection angle of light in KdS spacetime. In Sec.~\ref{comparing}, we present a detailed comparison between our results and those obtained in previous studies, and analyze potential observational candidates for detecting our corrections. Sec.~\ref{conclusion} offers a thorough summary. We adopt geometric unit ($G=c=1$) and the spacetime signature ($-,+,+,+$) throughout the paper.

\section{Three key aspects for light deflection in KdS spacetime}
\label{3problems}
The metric of KdS spacetime reads \cite{bardeen1973houches}
\begin{equation}
    \begin{aligned}
        \mathrm{d} s^{2} = & \frac{a^{2} \Delta _{\theta }\sin^{2} \theta -\Delta _{r}}{Q^{2} \Sigma }\mathrm{d} t^{2} +\frac{\Sigma }{\Delta _{r}}\mathrm{d} r^{2} +\frac{\Sigma }{\Delta _{\theta }}\mathrm{d} \theta ^{2}\\
         & +\frac{\sin^{2} \theta }{Q^{2} \Sigma }\left[ \Delta _{\theta }\left( a^{2} +r^{2}\right)^{2} -a^{2} \Delta _{r}\sin^{2} \theta \right]\mathrm{d} \phi ^{2}\\
         & -\frac{2a\sin^{2} \theta }{Q^{2} \Sigma }\left[ \Delta _{\theta }\left( a^{2} +r^{2}\right) -\Delta _{r}\right] \mathrm{d} t\mathrm{d} \phi ,
        \end{aligned}
    \label{kdsmetric0}
    \end{equation}
where $Q=1+\Lambda a^{2} /3$, $\Delta _{r} =\left( 1-\Lambda r^{2} /3\right)\left( r^{2} +a^{2}\right) -2Mr$, $\Sigma =r^{2} +a^{2}\cos^{2} \theta$, $\Delta _{\theta } =1+\Lambda a^{2}\cos^{2} \theta /3$.

\subsection{Orbit solution}
\label{Orbitsolution}
The original EOM for equatorial light (light confined to the equatorial plane) in KdS spacetime is a first-order differential equation, which reads \cite{sultana2013contribution}
\begin{equation}
    \begin{aligned}
        \left(\frac{\mathrm{d} u}{\mathrm{d} \phi }\right)^{2} = & \frac{1-b^{2} u^{2}}{b^{2}} +M\cdot 2u^{3} +\Lambda \cdot \frac{1}{3} -Ma\cdot \frac{4u}{b^{3}}\\
         & +a^{2} \cdot \frac{3u^{2} -2b^{2} u^{4}}{b^{2}} -a\Lambda \cdot \frac{2}{3b^{3} u^{2}} +\mathcal{O}\left( \epsilon ^{3}\right) ,
        \end{aligned}
        \label{dudphikds}
\end{equation}
where $u=1/r$, $\epsilon$ denotes the terms arising from combinations of $M/b$, $a/b$ and $\Lambda b^2$.

We first review the calculation of the orbit solution in Ref.~\cite{sultana2013contribution}. Differentiating Eq.~\eqref{dudphikds} with respect to $\phi$ yields a second-order differential equation 
\begin{equation}
        \frac{\mathrm{d}^{2} u}{\mathrm{d} \phi ^{2}} = -u+M\cdot 3u^{2} -Ma\cdot \frac{2}{b^{3}} +a^{2} \cdot \frac{3u-4u^{3} b^{2}}{b^{2}} + a \Lambda \cdot \frac{2}{3b^{3} u^{3}} +\mathcal{O}\left( \epsilon ^{3}\right) .
         \label{eomSultana}
\end{equation}
Assuming the orbit solution as $u=\sin/b+\delta u_1+\delta u_2$ ($\delta u_1$ and $\delta u_2$ are the first and second orders in $\epsilon$, respectively), and substituting it into Eq.~\eqref{eomSultana}, two differential equations are obtained
\begin{align}
    \frac{\mathrm{d}^{2}( \delta u_{1})}{\mathrm{d}\phi ^{2}} +\delta u_{1} &=3Mu_{0}^{2} , \label{deltau1}\\
    \frac{\mathrm{d}^{2}( \delta u_{2})}{\mathrm{d}\phi ^{2}} +\delta u_{2} &=6Mu_{0} u_{1} +\frac{3a^{2}}{b^{2}} u_{0} -4a^{2} u_{0}^{3} -\frac{2Ma}{b^{3}} +\frac{2a\Lambda }{3b^{3} u_{0}^{3}}. \label{deltau2}
\end{align}
The orbit solution is derived by solving Eqs.~\eqref{deltau1} and \eqref{deltau2}
\begin{equation}
\begin{aligned}
    u= & \frac{\sin \phi }{b} +M\cdot \frac{\cos (2\phi )+3}{2b^{2}} +M^{2} \cdot \frac{1}{16b^{3}}[ 30\pi \cos \phi \\
     & -3\sin (3\phi )-60\phi \cos \phi ] -Ma\cdot \frac{2}{b^{3}}\\
     & -a^{2} \cdot \frac{\sin (3\phi )}{8b^{3}} +a\Lambda \cdot \frac{\cos (2\phi )\csc \phi }{3} +\mathcal{O}\left( \epsilon ^{3}\right).
    \end{aligned}
    \label{uofphiSultana}
\end{equation}
In our opinion, Eq.~\eqref{uofphiSultana} does not fully satisfy the original EOM since Eq.~\eqref{dudphikds} can not be obtained based on Eq.~\eqref{uofphiSultana} (See \ref{SultanaOrbitSolution} for details). This concern is also raised in Refs.~\cite{bhadra2010gravitational,arakida2012effect,lake2002bending}.

Here, we present an orbit solution that exactly satisfies the original EOM by directly solving the original EOM itself. Assuming the orbit solution as $u=u_0+u_1+u_2$\footnote{The first-order orbit is sufficient for calculating the second-order deflection angle using the GW method. Here, we extend the orbit solution to second order solely for the purpose of comparison with previous results in Sec.~\ref{comparing}.} ($u_0$, $u_1$ and $u_2$ are respectively the zeroth, first and second orders in $\epsilon$), and substituting it into Eq.~\eqref{dudphikds} gives
\begin{equation}
    \begin{aligned}
        \left(\frac{\mathrm{d} u_{0}}{\mathrm{d} \phi }\right)^{2} = & \frac{1}{b^{2}} -u_{0}^{2}, \\
        2\frac{\mathrm{d} u_{0}}{\mathrm{d} \phi } \cdot \frac{\mathrm{d} u_{1}}{\mathrm{d} \phi } = & \frac{\Lambda }{3} +2Mu_{0}^{3} -2u_{0} u_{1}, \\
        2\frac{\mathrm{d} u_{0}}{\mathrm{d} \phi } \cdot \frac{\mathrm{d} u_{2}}{\mathrm{d} \phi } = & \frac{3a^{2} u_{0}^{2}}{b^{2}} -\frac{4Mau_{0}}{b^{3}} -\frac{2a\Lambda }{3b^{3} u_{0}^{2}} -\left(\frac{\mathrm{d} u_{1}}{\mathrm{d} \phi }\right)^{2}\\
         & -2a^{2} u_{0}^{4} +6Mu_{0}^{2} u_{1} -2u_{0} u_{2} -u_{1}^{2}.
        \end{aligned}
\end{equation}
Solving the above three equations yields
\begin{equation}
    \begin{aligned}
        u(\phi )= & \frac{\sin \phi }{b} +M\cdot \frac{\cos (2\phi )+3}{2b^{2}} +\Lambda \cdot \frac{b\sin \phi }{6} +M^{2} \cdot \frac{1}{16b^{3}}\\
         & \cdot [30\pi \cos \phi -3\sin (3\phi )-60\phi \cos \phi +37\sin \phi ]\\
         & -Ma\cdot \frac{2}{b^{3}} +M\Lambda \cdot \frac{\cos (2\phi )+3}{6} +a^{2} \cdot \frac{\sin^{3} \phi }{2b^{3}}\\
         & +a\Lambda \cdot \frac{\cos (2\phi )\csc \phi }{3} -\Lambda ^{2} \cdot \frac{b^{3}\sin \phi }{72} +\mathcal{O}\left( \epsilon ^{3}\right) ,
        \end{aligned}
        \label{uofphi01}
\end{equation}
where the integration constants are determined with the condition that $u$ reaches its maximum at $\phi=\pi/2$. To verify our orbit solution and support the calculations in Sec.~\ref{CalDefAngle}, we also derive the inverse solution of Eq.~\eqref{uofphi01}
\begin{equation}
    \phi(u) = \begin{cases}
    \Phi(u), & \text{if  } \left| \phi \right| <\frac{\pi}{2}, \\
    \pi - \Phi(u) , & \text{if  } \left| \phi \right| >\frac{\pi}{2},
        \end{cases}
        \label{phigamma}
\end{equation}
where
\begin{equation}
    \begin{aligned}
        \Phi (u)= & \arcsin (bu)+M\cdot \frac{b^{2} u^{2} -2}{b\sqrt{1-b^{2} u^{2}}} -\Lambda \cdot \frac{b^{3} u}{6\sqrt{1-b^{2} u^{2}}}\\
         & +M^{2} \cdot \left[\frac{u\left( 20b^{2} u^{2} -3b^{4} u^{4} -15\right)}{4b\left( 1-b^{2} u^{2}\right)^{3/2}}\right. \\
         & \left. -\frac{15\arccos (bu)}{4b^{2}}\right] +Ma\cdot \frac{2}{b^{2}\sqrt{1-b^{2} u^{2}}}\\
         & +M\Lambda \cdot \frac{b\left( 3b^{2} u^{2} -2\right)}{6\left( 1-b^{2} u^{2}\right)^{3/2}} -a^{2} \cdot \frac{bu^{3}}{2\sqrt{1-b^{2} u^{2}}}\\
         & +a\Lambda \cdot \frac{\left( 2b^{2} u^{2} -1\right)}{3u\sqrt{1-b^{2} u^{2}}} +\Lambda ^{2} \cdot \frac{b^{5} u\left( 3-2b^{2} u^{2}\right)}{72\left( 1-b^{2} u^{2}\right)^{3/2}}\\
         & +\mathcal{O}\left( \epsilon ^{3}\right) .
        \end{aligned}
        \label{PHIofu}
\end{equation}
For the first-order derivative of Eq.~\eqref{uofphi01} with respect to $\phi$, substituting Eq.~\eqref{phigamma} into it and then squaring the result brings about a formula that is entirely consistent with the original EOM, which validates the correctness and precision of our orbit solution.

A brief discussion on the discrepancy between Eqs.~\eqref{uofphiSultana} and \eqref{uofphi01} is presented in \ref{discrepancy}.

\subsection{Positions of the source and observer}
\label{Positionsofsourceandobserver}
The KdS spacetime features a cosmological horizon at approximately $\sqrt{3/\Lambda}$ which the source and observer cannot extend beyond. Therefore, the assumption that the source and observer are located at infinity is invalid for KdS spacetime.

In Ref.~\cite{sultana2013contribution}, the positions of the source and observer are determined with the approximation that the azimuthal coordinate is "sufficiently small" ($\phi_0\ll 1$). Sultana believes this approximation is valid because it ensures that all significant bending from the massive object has been accounted for, specifically, the Einstein deflection angle (the deflection angle of equatorial light in Kerr spacetime) $\delta_{Kerr}=4M/b+15\pi M^2/4b^2-4Ma/b^2$ has been achieved. Consequently, he concluded that $\phi_0\ll 1$ ensures the source and observer are located within a finite region.

\begin{figure}[!ht]
    \centering
    \includegraphics[width=1\columnwidth]{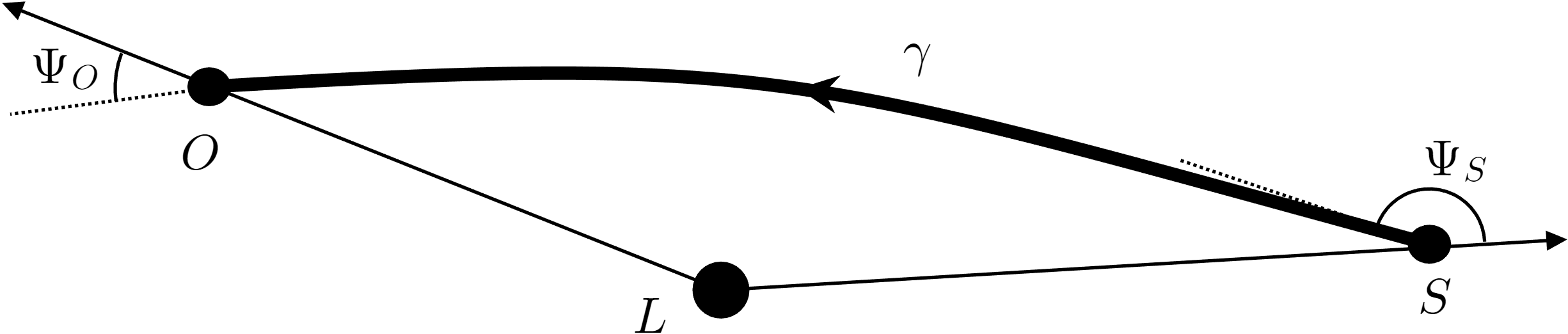}
    \caption{The schematic for the finite-distance deflection angle.}
    \label{fig-1}
  \end{figure}

In this paper, the finite-distance deflection angle which incorporates the distances of the source and observer to the lens is adopted to overcome the challenges posed by the cosmological horizon. In 2017, Ishihara \textit{et al.} introduced this concept for light in curved spacetimes \cite{ishihara2016gravitational}. As shown in Fig.~\ref{fig-1}, 
$L$ is the lens, $\gamma$ is the trajectory of the light from source $S$ to observer $O$. $\Psi_S$ and $\Psi_O$ represent the angles between the outward radial direction and the tangent along $\gamma$ at $S$ and $O$, respectively. The finite-distance deflection angle is defined as \cite{ishihara2016gravitational}
\begin{equation}
    \delta = \Psi_O- \Psi_S + \phi_{OS},
    \label{defineangle}
\end{equation}
where $\phi_{OS}=\phi_O-\phi_S$, with $\phi_O$ and $\phi_S$ being the azimuthal coordinates of $O$ and $S$, respectively. This definition has been proved geometrically invariant, i.e., well-defined, with the GW method \cite{ishihara2016gravitational}, and has been widely used in studying the deflection of massless and massive particles in various spacetimes \cite{li2020thefinitedistance,takizawa2020gravitational,ishihara2017finite,ono2018deflection,kumar2019shadow,li2020finite,li2021kerr,li2021deflection,huang2022generalized,huang2023extending,huang2023finite,ono2017gravitomagnetic,ono2019deflection,haroon2019shadow,crisnejo2019finite,ono2019effects,li2020circular,belhaj2022light,li2021kerrnewman,li2022deflection,pantig2022testing}. When the source and the observer approach infinity (i.e., $\Psi_S\to \pi$ and $\Psi_O \to 0$), Eq.~\eqref{defineangle} reduces to the conventional infinite-distance deflection angle $\phi_{OS}-\pi$. In this paper, Eq.~\eqref{defineangle} is applied to KdS spacetime, providing an alternative approach to addressing the position issue of the source and observer.

\subsection{Staticity of the source and observer}
\label{Staticityofsourceandobserver}
As stated in Ref.~\cite{sultana2013contribution}, the reduced space used to discuss the light deflection is derived with $\mathrm{d}t=0$, indicating that the source and observer are assumed to be at rest in a static slice of spacetime. Such staticity issue is also raised in Refs.~\cite{ishihara2016gravitational,park2008rigorous,simpson2010lensing}.

Following the GW method, we adopt the Randers optical space in this paper, which may be an appropriate space for discussing the light in KdS spacetime. The geodesic of the Randers optical space acts as the spatial projection of null geodesics in stationary axially symmetric (SAS) spacetimes, which is guaranteed by Fermat's principle \cite{werner2012gravitational}. Thus, analyzing light deflection via geodesics in such space eliminates the assumption that the source and observer reside at rest in a static slice of spacetime. The metric of SAS spacetimes can be written as
\begin{equation}
    \mathrm{d} s^{2}  = g_{tt}\mathrm{d} t^{2} +2g_{t\phi }\mathrm{d} t\mathrm{d} \phi +g_{rr}\mathrm{d} r^{2} +g_{\theta \theta }\mathrm{d} \theta ^{2} +g_{\phi \phi }\mathrm{d} \phi ^{2}.
    \label{SASMetric}
\end{equation}
The Randers optical space ($M^{(RO)}$) for an SAS spacetime is defined by the following metric 
\begin{equation}
    \mathrm{d} \tilde{l}  = \sqrt{\alpha_{ij}\mathrm{d}x^i\mathrm{d}x^j} +\beta_i \mathrm{d}x^i,
    \label{opticalRandersSAS}
  \end{equation}
  where
\begin{align}
     \alpha_{ij}\mathrm{d}x^i \mathrm{d}x^j 
     =&  \frac{g_{rr}}{-g_{tt}}\mathrm{d} r^{2} +\frac{g_{\theta\theta}}{-g_{tt}}\mathrm{d} \theta ^{2} +\frac{g_{t\phi}^{2}-g_{tt} g_{\phi\phi}}{g_{tt}^{2}}\mathrm{d} \phi ^{2} , \label{alphaijSAS} \\
     \beta_i \mathrm{d}x^i = & \frac{g_{t\phi}}{-g_{tt}} \mathrm{d} \phi. \label{betaSAS}
\end{align}
The light deflection in an SAS spacetime can be studied in the corresponding $M^{(RO)}$.

We designate the three-dimensional space determined by Eq.~\eqref{alphaijSAS} as $M^{(\alpha 3)}$. A geodesic in $M^{(RO)}$ can be put in one-to-one correspondence with a curve, denoted by $\gamma$, in $M^{(\alpha 3)}$ \cite{ono2017gravitomagnetic}. Therefore, the light deflection in an SAS spacetime can also be studied in $M^{(\alpha 3)}$ \cite{ono2017gravitomagnetic}. Specifically, for the equatorial light in KdS spacetime, the corresponding $M^{(\alpha 3)}$ reduces to a two dimensional space (denoted by $M^{(\alpha 2)}_{kds}$). Substituting $\theta=\pi/2$, $\mathrm{d}\theta=0$ and Eq.\eqref{kdsmetric0} into Eq.~\eqref{opticalRandersSAS} yields the metric of $M^{(\alpha 2)}_{kds}$, i.e., $\mathrm{d}l^2 = \alpha_{rr}\mathrm{d}r^2 + \alpha_{\phi\phi}\mathrm{d}\phi^2$ where
\begin{equation}
    \begin{aligned}
        \alpha _{rr} = & \frac{r^{3}\left( a^{2} \Lambda +3\right)^{2}}{\mathcal{Z}\left( r\mathcal{Z} -3a^{2}\right)} , \quad
        \alpha _{\phi \phi } =  \frac{3r^{2}\left( 3a^{2} -r\mathcal{Z}\right)}{\mathcal{Z}^{2}},
        \end{aligned}
        \label{JMRFkds}
\end{equation}
and the corresponding one-form 
\begin{equation}
    \beta _{\phi } = a(\mathcal{Z} +3r)/\mathcal{Z}, \label{betaphi}    
\end{equation}
with $\mathcal{Z} =\Lambda r^{3} +\left( \Lambda a^{2} -3\right) r+6M$.

\section{Deflection angle by generalized GW method} \label{CalDefAngle}
\label{Calculation}
In this section, using the generalized GW method for stationary spacetimes \cite{huang2024generalized}, we calculate the deflection angle of light in KdS spacetime based on the orbit solution in Eq.~\eqref{uofphi01}, the definition of the finite-distance deflection angle in Eq.~\eqref{defineangle}, and the two-dimensional space $M^{(\alpha 2)}_{kds}$ in Eq.~\eqref{JMRFkds}.

\begin{figure}[!ht]
    \centering
    \includegraphics[width=1\columnwidth]{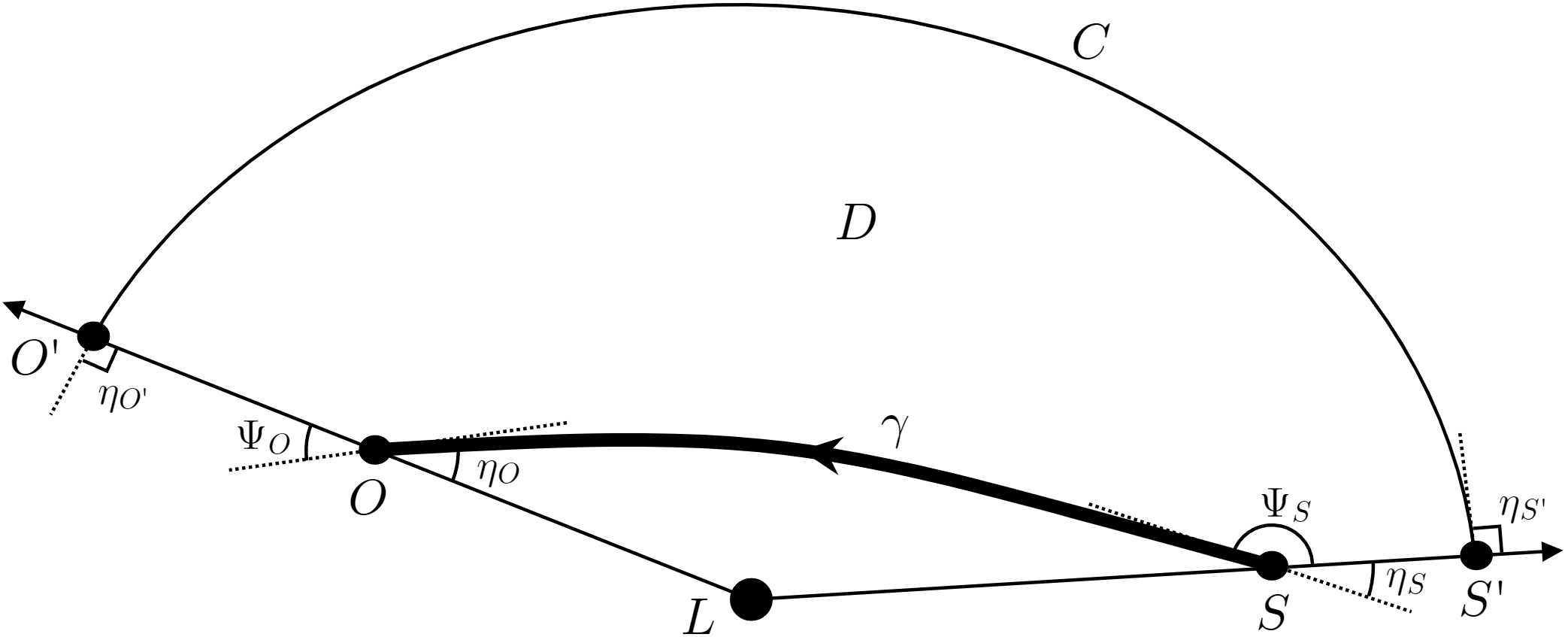}
    \caption{The quadrilateral region $D=^{O'}_{O}\square^{S'}_S$ in $M^{(\alpha 2)}_{kds}$.}
    \label{fig-2}
  \end{figure}

As shown in Fig.~\ref{fig-2},
in $M^{(\alpha 2)}_{kds}$, $\gamma$, $S$, $O$, $L$, $\Psi_S$ and $\Psi_O$ carry the same meaning as those illustrated in Fig.~\ref{fig-1}. $C=\overset{\curvearrowright}{S' O'}$ is an auxiliary circular arc with $r=r_c\in (r_\gamma^{max}, r_\Lambda)$, where $r_\gamma^{max}$ represents the maximal radial coordinate of $\gamma$ and $r_\Lambda$ denotes the cosmological horizon of KdS spacetime. $C$ intersects with the outward radial curves $\overrightarrow{LS}$ and $\overrightarrow{LO}$ at $S'$ and $O'$, respectively. $\eta_S$, $\eta_{S'}$, $\eta_{O'}$ and $\eta_O$ indicate the jump angles at $S$, $S'$, $O'$ and $O$, in that order. Then we get a quadrilateral region $D=^{O'}_{O}\square^{S'}_S$, which is surrounded by $\gamma$, $\overrightarrow{SS'}$, $C$, and $\overrightarrow{OO'}$ . 

Applying the Gauss-Bonnet theorem (pp. 272 and 277 in Ref.~\cite{manfredo1976carmo}) to $D$ brings about
\begin{equation}
    \begin{aligned}
        &\iint _{D} K\mathrm{d} S+\int _{\overrightarrow{SS} '} \kappa \mathrm{d} l+\int _{C} \kappa \mathrm{d} l+\int _{\overrightarrow{O' O}} \kappa \mathrm{d} l\\
        &+\int _{\overset{\curvearrowright }{OS}} \kappa \mathrm{d} l +\eta _{S} +\eta _{S'} +\eta _{O'} +\eta _{O} =2\pi \chi (D ),
    \end{aligned}
    \label{gbD}
  \end{equation}
where $K$, $\mathrm{d}S$ and $\chi\left(D\right)$ stand for the Gaussian curvature, area element and Euler characteristic number of $D$, respectively; $\kappa$ and $\mathrm{d}l$ denote the geodesic curvature and line element of curves, respectively. Substituting $\kappa(\overrightarrow{SS'})=\kappa(\overrightarrow{O'O})=0$ (refer to Appendix A of Ref.~\cite{huang2023finite} for the proof), $\int_{\overset{\curvearrowright}{OS}} \kappa \mathrm{d}l=-\int_\gamma \kappa \mathrm{d}l$, $\eta_S = \pi - \Psi_S$, $\eta_{S'}=\eta_{O'}=\pi/2$, $\eta_O=\Psi_O$, and $\chi\left(D\right)=1$ ($D$ is simply connected) into Eq.~\eqref{gbD} and combining the result with the definition Eq.~\eqref{defineangle}, the finite-distance deflection angle can be expressed as
\begin{equation}
    \delta = - \iint_{D} K \mathrm{d}S -\int_{C} \kappa \mathrm{d}l   +\phi_{OS}+ \int_\gamma \kappa \mathrm{d}l.
    \label{deltageneral}
\end{equation}
According to the generalized GW method for rotating spacetimes (Secs.~4.1 and 4.4 of Ref.~\cite{huang2024generalized}), the above formula becomes
\begin{equation}
    \delta = \int_{\phi_S}^{\phi_O} f(r_\gamma) \mathrm{d}\phi,
    \label{delta0}
\end{equation}
where $r_\gamma$ represents the radial coordinate of $\gamma$ and is expressed in terms of the azimuthal coordinate $\phi$ in calculation, $f$ is defined by
\begin{equation}
    f(r)=1-\frac{\alpha_{\phi\phi,r}}{2\sqrt{\alpha_{rr}\alpha_{\phi\phi}}} -\beta_{\phi,r} \sqrt{ \frac{r^4}{\alpha_{\phi\phi}} \left( \frac{\mathrm{d}u}{\mathrm{d}\phi}\right)^2+ \frac{1}{\alpha_{rr}} }. 
    \label{fr}
\end{equation}

Substituting the metric of $M^{(\alpha 2)}_{kds}$ in Eq.~\eqref{JMRFkds}, the corresponding one-form in Eq.~\eqref{betaphi}, and $\left(\mathrm{d}u/\mathrm{d}\phi\right)^2$ in Eq.~\eqref{dudphikds} into Eq.~\eqref{fr}, then combing the result with our orbit solution $r_\gamma$ (the reciprocal of Eq.~\eqref{uofphi01}), we have
\begin{equation}
    \begin{aligned}
        f(r_{\gamma } )= & M\cdot \frac{2\sin \phi }{b} -\Lambda \cdot \frac{b^{2}\csc^{2} \phi }{6} +M^{2} \cdot \frac{1}{4b^{2}}\Bigl[ 15\\
         & +\cos (2\phi )\Bigr] -Ma\cdot \frac{2\sin \phi }{b^{2}} +M\Lambda \cdot \frac{b\sin \phi }{3}\\
         & \cdot \Bigl( 1+2\csc^{4} \phi -\csc^{2} \phi \Bigr) +a\Lambda \cdot \frac{2b\csc^{2} \phi }{3}\\
         & -\Lambda ^{2} \cdot \frac{b^{4}\csc^{4} \phi }{72}\Bigl[ 2\cos (2\phi )+1\Bigr] +\mathcal{O}\left( \epsilon ^{3}\right) .
        \end{aligned}
        \label{frgamma}
\end{equation}
It should be noted that Eq.~\eqref{frgamma}, as well as the final deflection angle, can be derived using only the first-order of the orbit solution, i.e., $u(\phi)=\sin\phi/b+M(\cos^2\phi+1)/b^2+\Lambda b \sin\phi/6$, owing to the inherent characteristics of the GW method. Denoting the indefinite integral of $f(r_\gamma)$ as $F(\phi)$, Eq.~\eqref{delta0} can be recast as $\delta = F\left(\phi_O\right)-F\left(\phi_S\right)$, namely $\delta=F\left[\pi-\Phi\left(u_O\right)\right]-F\left[\Phi\left(u_S\right)\right]$, where Eq.~\eqref{phigamma} is used and $\phi_S<\pi/2<\phi_O$ is assumed. Finally, with the expression of $\Phi(u)$ in Eq.~\eqref{PHIofu}, we obtain the finite-distance deflection angle of equatorial light in KdS spacetime
\begin{equation}
    \begin{aligned}
        \delta = & M\cdot \frac{2}{b}\left(\sqrt{1-b^{2} u_{O}^{2}} +\sqrt{1-b^{2} u_{S}^{2}}\right)\\
         & -\Lambda \cdot \frac{b}{6}\left(\frac{\sqrt{1-b^{2} u_{O}^{2}}}{u_{O}} +\frac{\sqrt{1-b^{2} u_{S}^{2}}}{u_{S}}\right)\\
         & +M^{2} \cdot \frac{1}{4b^{2}}\Biggl\{15\bigl[\arccos (bu_{O} )+\arccos (bu_{S} )\bigr]\\
         & +\frac{15bu_{O} -7b^{3} u_{O}^{3}}{\sqrt{1-b^{2} u_{O}^{2}}} +\frac{15bu_{S} -7b^{3} u_{S}^{3}}{\sqrt{1-b^{2} u_{S}^{2}}}\Biggr\}\\
         & -Ma\cdot \frac{2}{b^{2}}\left(\sqrt{1-b^{2} u_{O}^{2}} +\sqrt{1-b^{2} u_{S}^{2}}\right)\\
         & +M\Lambda \cdot \frac{b}{6}\left(\frac{1}{\sqrt{1-b^{2} u_{O}^{2}}} +\frac{1}{\sqrt{1-b^{2} u_{S}^{2}}}\right)\\
         & +a\Lambda \cdot \frac{2}{3}\left(\frac{\sqrt{1-b^{2} u_{O}^{2}}}{u_{O}} +\frac{\sqrt{1-b^{2} u_{S}^{2}}}{u_{S}}\right)\\
         & -\Lambda ^{2} \cdot \frac{b}{72}\left(\frac{1-b^{2} u_{O}^{2} -2b^{4} u_{O}^{4}}{u_{O}^{3}\sqrt{1-b^{2} u_{O}^{2}}} +\frac{1-b^{2} u_{S}^{2} -2b^{4} u_{S}^{4}}{u_{S}^{3}\sqrt{1-b^{2} u_{S}^{2}}}\right)\\
         & +\mathcal{O}\left( \epsilon ^{3}\right).
        \end{aligned}
        \label{deltarst}
\end{equation}

\section{Discussions}
\label{comparing}
\subsection{Comparing with previous results}
We perform a term-by-term comparison between Eq.~\eqref{deltarst} and the previous results derived for Schwarzschild-de Sitter (SdS) and KdS spacetimes using various approaches.

\subsubsection{$M$, $M^2$, and $Ma$ terms}
Li calculated the finite-distance deflection angle of particles in Kerr spacetime, as shown in Eq.~(42) of Ref.~\cite{li2020thefinitedistance}. For massless particles ($v=1$), the $M$, $M^2$, and $Ma$ terms therein are consistent with those in Eq.~\eqref{deltarst}.

Under the infinite-distance limit of the source and observer ($u_S=u_O=0$), the $M$, $M^2$, and $Ma$ terms in Eq.~\eqref{deltarst} simplify to $4M/b$, $15\pi M^2/(4b^2)$, and $-4Ma/b^2$, respectively. These results align with those in Eq.~\eqref{DAKdSSultana}, which is obtained assuming the source and observer are very far from the lens (far-distance approximation).

\subsubsection{$\Lambda$ term}
The linear term of $\Lambda$ in Eq.~\eqref{deltarst} is consistent with those in Refs.~\cite{ishihara2016gravitational} (Eq.~(37)) and \cite{takizawa2020gravitational} (Eq.~(23)), which are committed to calculating the finite-distance deflection angle of light in SdS spacetime.

Sereno calculated the deflection angle of light in SdS spacetime under the far-distance approximation \cite{sereno2008influence}. The $\Lambda$ term in his result is $\Lambda b^3(u_O+u_S)/6$, which seems to differ from ours. This discrepancy arises because he only considers the increment of the azimuthal coordinate, namely $\delta_{Sereno}=\phi_{OS}=\phi_O-\phi_S$. In fact, using Eq.~\eqref{phigamma}, we can drive the $\Lambda$ term for $\phi_{OS}$ as $\left( \Lambda b^{3} /6\right)\left( u_{O} /\sqrt{1-b^{2} u_{O}^{2}} +u_{S} /\sqrt{1-b^{2} u_{S}^{2}}\right)$ which agrees with that in Ref.~\cite{sereno2008influence} under the far-distance approximation ($b^2u_O^2\sim0$).

An interesting observation is that in Refs.~\cite{arakida2012effect,arakida2018light,takizawa2022gravitational}, which focus on the deflection angle of light in SdS spacetime, the $\Lambda$ term disappears. In fact, the deflection angle in these three papers are calculated by considering the deviation between the trajectory in SdS spacetime and that in de Sitter spacetime, thus the contribution arising solely from $\Lambda$ disappears (see Appendix \ref{desitterbackground} for more details). While in this paper and Refs.\cite{sereno2008influence,sereno2009role,sultana2013contribution,ishihara2016gravitational,takizawa2020gravitational}, the deflection angle is defined as the deviation between the trajectory in SdS spacetime and that in Minkowski spacetime. It can be verified that, within our framework, subtracting the deflection angle of de Sitter spacetime from that of SdS spacetime leads to the disappearance of the $\Lambda$ term.

It seems difficult to precisely identify the cause of the difference between the $\Lambda$ term in Eqs.~\eqref{DAKdSSultana} and ~\eqref{deltarst}. Additionally, explaining the discrepancy between the $\Lambda$ term in Eq.~\eqref{DAKdSSultana} and that in results reported by others is equally challenging.

\subsubsection{$M\Lambda$ term}
The $M\Lambda$ term in Eq.~\eqref{deltarst} is consistent with those in Refs.~\cite{ishihara2016gravitational} (Eq.~37), \cite{takizawa2020gravitational} (Eq.~(23)) and \cite{takizawa2022gravitational} (Eq.~(79)), which are committed to calculating the finite-distance deflection angle of light in SdS spacetime.

Under the infinite-distance approximation, the $M\Lambda$ term in Eq.~\eqref{deltarst} becomes $M\Lambda b/3$, which is the half of the results in Refs.~\cite{sereno2008influence} (Eq.~(5)), \cite{bhadra2010gravitational} (Eq.~(15)), and \cite{arakida2012effect} (Eq.~(14)). This discrepancy arises because only the the increment of the azimuthal coordinate, namely $\phi_{OS}=\phi_O-\phi_S$, is considered as the deflection angle in Refs.~\cite{sereno2008influence,bhadra2010gravitational,arakida2012effect}. Using Eq.~\eqref{phigamma}, we can obtain the $M\Lambda$ term of the $\phi_{OS}$ as
\begin{equation}
    M\Lambda \cdot \frac{b}{6} \left[\frac{2-3b^2 u_O^2}{(1-b^2 u_O^2)^{3/2}}+\frac{2-3b^2 u_S^2}{(1-b^2 u_S^2)^{3/2}}\right].
    \label{MLambda}
\end{equation}
When $u_S=u_O=0$, Eq.~\eqref{MLambda} simplifies to $2M\Lambda b/3$, which is consistent with the results in Refs.~\cite{sereno2008influence,bhadra2010gravitational,arakida2012effect}.

It seems difficult to precisely identify the cause of the difference between the $M\Lambda$ term in Eqs.~\eqref{DAKdSSultana} and ~\eqref{deltarst}. Additionally, explaining the discrepancy between the $M\Lambda$ term in Eq.~\eqref{DAKdSSultana} and that in results reported by others is equally challenging.

\subsubsection{$a\Lambda$ term}
Currently, only this paper and Ref.~\cite{sultana2013contribution} have provided the $a\Lambda$ term. However, the two results differ, and it is challenging to pinpoint the underlying reason for the divergence. Both results have the same sign, i.e., positive. Therefore, for photons with angular momentum aligned with the spin of the central body (prograde light), the deflection angle will increase due to the $a\Lambda$ term. Conversely, for retrograde light, the deflection angle will decrease.

\subsubsection{$\Lambda^2$ term}
This paper is the first to present the $\Lambda^2$ term in the deflection angle of light for spacetimes involving the cosmological constant. Compared with Eq. \eqref{DAKdSSultana}, the $\Lambda^2$ term in Eq.\eqref{deltarst} comes from directly solving the original EOM, which guarantees the $\Lambda$ and $\Lambda^2$ terms are retained in the orbit solution.

Similar to the linear term of $\Lambda$, the $\Lambda^2$ term also reduces the deflection angle, which is consistent with the intuition that the cosmological constant exerts repulsive effects.

\subsection{Possible Observational Candidates}
\label{observation}
The current astrometric space mission Gaia \cite{gaia,brown2025gaia}, as well as the upcoming Japanese mission JASMINE (Japan Astrometry Satellite Mission for Infrared Exploration) \cite{jasmine}, are expected to achieve an angular precision of nearly 10 $\mu$as. This level of accuracy may provide observational support for the correction we propose. (In the following discussion, our correction relative to Sultana's result is denoted as $\Delta$, namely, $\Delta= \left| Eq.~\eqref{deltarst} - Eq.~\eqref{DAKdSSultana} \right|$.)

\begin{figure}[!ht]
    \centering
    \includegraphics[width=1\columnwidth]{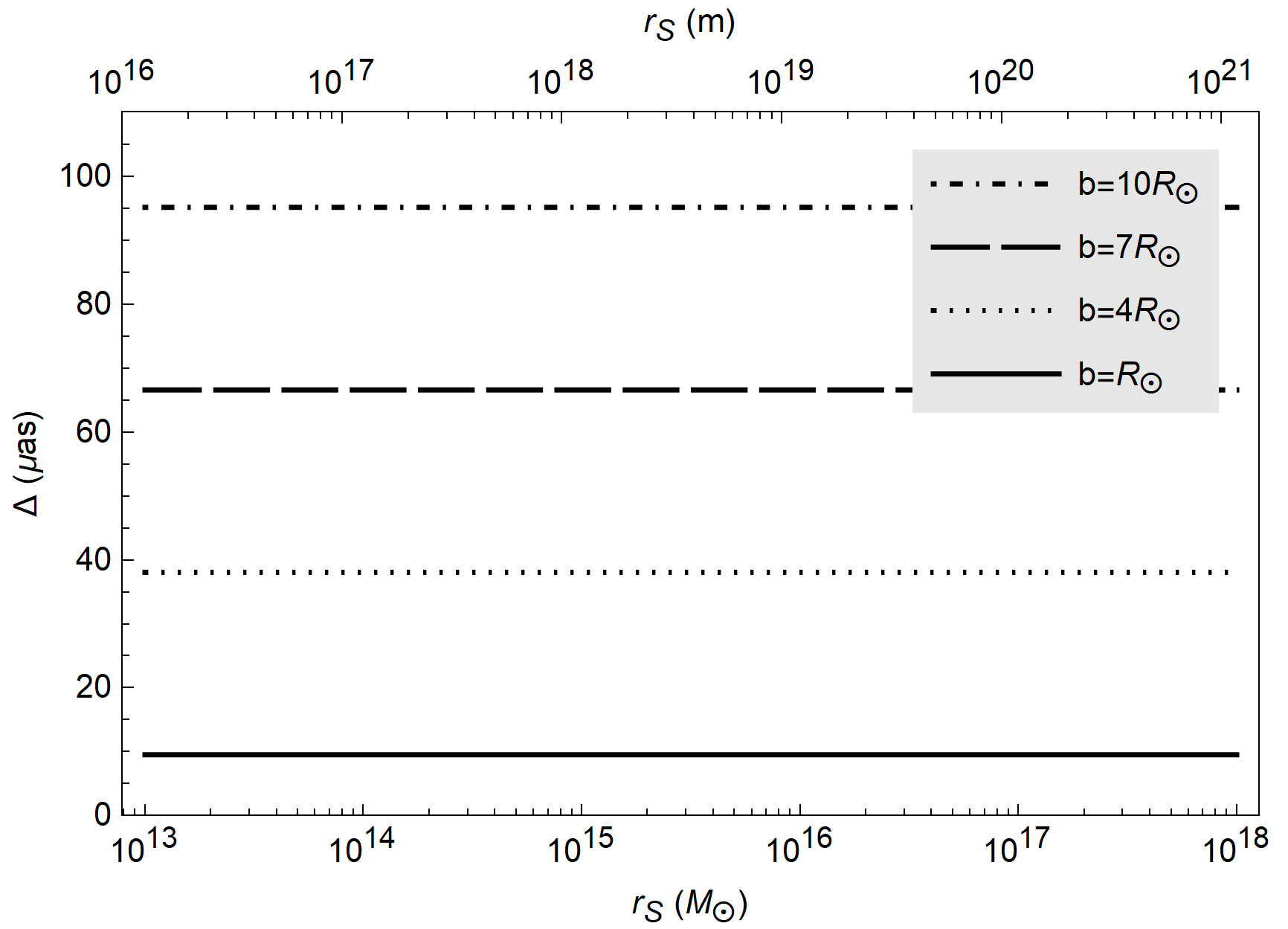}
    \caption{The $\Delta$ for the case where the Sun acts as the lens object. We take $10^{13}M_\odot$ (in geometric units, $M_\odot$ is the mass of the Sun) as the lower bound for $r_S$, since the closest star to the Sun (Proxima Centauri) located at a distance of approximately $10^{16}$m (or $10^{13}M_\odot$) \cite{anglada2016terrestrial}. The impact parameter $b$ is considered at five values: $R_\odot$, $2R_\odot$, $3R_\odot$, $4R_\odot$, and $5R_\odot$, where $b=R_\odot$ represents a grazing light ray—that is, a light ray passing just above the solar surface.}
    \label{fig-3}
  \end{figure}

Fig.~\ref{fig-3} illustrates the behavior of $\Delta$ for the Sun as the lensing object.
While the distance between the Sun and the source appears to have no significant effect on $\Delta$ in this case, the impact parameter does play a crucial role. Specifically, $\Delta$ increases with the impact parameter, and even at its smallest value (the radial of the Sun), the corresponding $\Delta$ approaches 10$\mu$as. Thus, such corrections might be reachable in the near-future astronomical observations.

\begin{figure}[!ht]
    \centering
    \includegraphics[width=1\columnwidth]{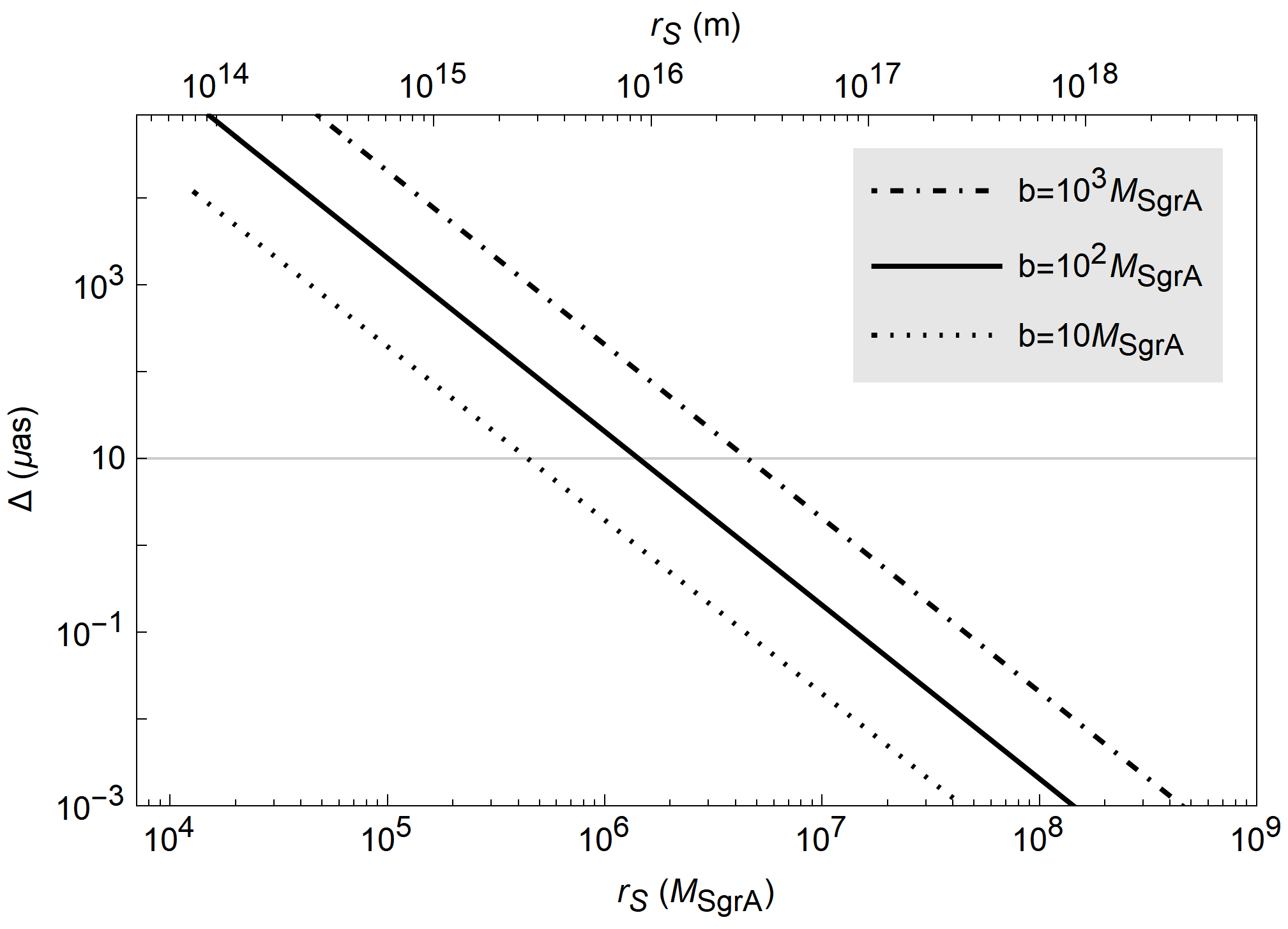}
    \caption{The $\Delta$ for the case where Sgr A* acts as the lens object. We adopt $10^4M_{SgrA}$ (in geometric units, where $M_{SgrA}$ denotes the mass of Sgr A*) as the lower limit for $r_S$, based on the fact that the currently known closest star to Sgr A* is located at an average distance of approximately $10^{14}$m (or $10^4M_{SgrA}$) \cite{peissker2020s62on,mychelkin2024weak}. The impact parameter $b$ is considered at three representative values: $10M_{SgrA}$, $10^2M_{SgrA}$, and $10^3M_{SgrA}$. The values for the mass and spin of Sgr A* utilized in our study are sourced from Ref.~\cite{fakiyama2022first}.}
     \label{fig-4}
  \end{figure}

Fig.~\ref{fig-4}
presents the results for the case where Sgr A* acts as the lens object. Here, $\Delta$ decreases with increasing source distance. Similar to the case of the Sun, $\Delta$ also increases with the impact parameter. The results indicate that $\Delta$ can reach the order of 10$\mu$as (or more) when the source stars are sufficiently close to Sgr A*. For such stars, the corrections proposed in this paper become observationally relevant. Given current technological advancements, detecting $\Delta$ for stars near Sgr A* may become feasible in the near future.

\section{Conclusion}
\label{conclusion}
The contribution of the cosmological constant to the deflection of light in rotating spacetimes, specifically KdS spacetime, was previously explored by Sultana. In this paper, by using the generalized GW method, we re-examine and refine the result from three aspects---orbit solution, positions of the source and observer, and staticity of both. Our result updates all terms involving cosmological constant, i.e., the $\Lambda$, $M\Lambda$, $a\Lambda$, and $\Lambda^2$ terms, thereby improving upon earlier findings. It should be noted that the $\Lambda$ and $M\Lambda$ terms have also been discussed by other researchers in the context of SdS spacetime. Those results show apparent discrepancies with our formulas, which can be attributed to our more general calculation conditions. To test our refined predictions, we consider lensing scenarios involving the Sun and Sgr A*. Given the increasing precision of astronomical observations, our correction may be verifiable with near-future observational data.

\section*{Acknowledgments}
This work was supported in part by Natural Science Foundation of Hunan Province for Youths Grant No. 2024JJ6210 and 2024JJ6211, the National Natural Science Foundation of China Grant No. 12405054, 12375045, 12205093, and 12405053, and the Science Research Fund of Hunan Provincial Education Department No. 24B0476, 21A0297 and 24C0229.

\appendix
\section{Sultana's orbit solution}
\label{SultanaOrbitSolution}
From where we stand, although Eq.~\eqref{uofphiSultana} satisfies the second-order differential equation~\eqref{eomSultana}, it is not a good solution to the original EOM Eq.~\eqref{dudphikds}. Below is our demonstration: Firstly, by differentiating Eq.~\eqref{uofphiSultana} with respect to $\phi$ and squaring the result, we get
\begin{equation}
    \begin{aligned}
        \left(\frac{\mathrm{d} u}{\mathrm{d} \phi }\right)^{2} = & \frac{\cos^{2} \phi }{b^{2}} -M\cdot \frac{2\sin (2\phi )\cos \phi }{b^{3}}-\\
         & M^{2} \cdot \frac{\cos \phi }{8b^{4}} \Big[ 30(\pi -2\phi )\sin \phi +52\cos \phi \\
         & +17\cos (3\phi )\Big] -a^{2} \cdot \frac{3\cos (3\phi )\cos \phi }{4b^{4}}\\
         & +a\Lambda \cdot \frac{2[\cos (2\phi )-2]\cot^{2} \phi }{3b} +\mathcal{O}\left( \epsilon ^{3}\right).
        \end{aligned}
        \label{dudphiSultanaTest}
\end{equation}
Secondly, we derive the inverse solution of Eq.~\eqref{uofphiSultana}
\begin{equation}
    \begin{aligned}
        \phi = & \arcsin (bu)+M\cdot \frac{b^{2} u^{2} -2}{b\sqrt{1-b^{2} u^{2}}}\\
         & +M^{2} \cdot \left[\frac{bu^{3}}{2\left( 1-b^{2} u^{2}\right)^{3/2}} +\frac{3bu^{3}}{4\sqrt{1-b^{2} u^{2}}}\right. \\
         & \left. -\frac{23u}{16b\sqrt{1-b^{2} u^{2}}} -\frac{15\arccos( bu)}{4b^{2}}\right]\\
         & +Ma\cdot \frac{2}{b^{2}\sqrt{1-b^{2} u^{2}}} -a^{2} \cdot \frac{u\left( 4b^{2} u^{2} -3\right)}{8b\sqrt{1-b^{2} u^{2}}}\\
         & +a\Lambda \cdot \frac{2b^{2} u^{2} -1}{3u\sqrt{1-b^{2} u^{2}}} +\mathcal{O}\left( \epsilon ^{3}\right) , \ \ \ \left(\left|\phi\right|<\frac{\pi}{2}\right).
        \end{aligned}
        \label{phiofuSultana}
\end{equation}
Substituting Eq.~\eqref{phiofuSultana} into the $\mathrm{d}^2u/\mathrm{d}\phi^2$, which is derived by differentiating Eq.~\eqref{uofphiSultana} with respect to $\phi$, we can obtain a formula that is completely consistent with Eq.~\eqref{eomSultana}. However, substituting Eq.~\eqref{phiofuSultana} into Eq.~\eqref{dudphiSultanaTest} leads to
\begin{equation}
    \begin{aligned}
        \left(\frac{\mathrm{d} u}{\mathrm{d} \phi }\right)^{2} = & \frac{1-b^{2} u^{2}}{b^{2}} +M\cdot 2u^{3} -M^{2} \cdot \frac{37}{8b^{4}}-\\
         & Ma\cdot \frac{4u}{b^{3}} +a^{2}\cdot\left(\frac{3u^{2} -2b^{2} u^{4}}{b^{2}} -\frac{3}{4b^{4}}\right)\\
         & -a\Lambda \cdot \frac{2}{3b^{3} u^{2}} + \mathcal{O}\left(\epsilon^3 \right),
        \end{aligned}
        \label{dudphi2sultana}
\end{equation}
which is different from the original EOM Eq.~\eqref{dudphikds} in $\Lambda$, $M^2$ and $a^2$ terms. Therefore, the orbit solution Eq.~\eqref{uofphiSultana} does not fully satisfies the original EOM.

\section{Discrepancies in orbit solutions}
\label{discrepancy}
The discrepancy between Eqs.~\eqref{uofphiSultana} and \eqref{uofphi01} can be summarized in two aspects: 
\begin{itemize} 
    \item  The $\Lambda$, $M\Lambda$ and $\Lambda^2$ terms appear in Eq.~\eqref{uofphi01} but are absent in Eq.~\eqref{uofphiSultana}. This suggests that differentiating Eq.~\eqref{dudphikds} eliminates the $\Lambda$ term, resulting in the disappearance of the associated $\Lambda$, $M\Lambda$ and $\Lambda^2$ terms in the derived differential equations (Eqs.~\eqref{deltau1} and \eqref{deltau2}) and orbit solution (Eq.~\eqref{uofphiSultana}). As for the $a\Lambda$ term present in Eqs.~\eqref{deltau2} and \eqref{uofphiSultana}, it originates solely from the $a\Lambda$ term in Eq.~\eqref{eomSultana}, since the $a$ term vanishes in both Eqs.~\eqref{dudphikds} and \eqref{eomSultana}. Additionally, we note that the $(\mathrm{d}u/\mathrm{d}\phi)^2$ derived from Eq.~\eqref{uofphiSultana}, i.e., Eq.~\eqref{dudphi2sultana}, losts the $\Lambda$ term compared to the original EOM. 
    \item  The coefficients of the $M^2$ and $a^2$ terms in Eq.~\eqref{uofphiSultana} differ from those in Eq.~\eqref{uofphi01}. Following the calculation procedure in Ref.~\cite{sultana2013contribution}, we obtain the $a^2$ term as $-a^2 \sin(3\phi)/(8b^2)$, which is consistent with that in Eq.~\eqref{uofphiSultana}, whereas the $M^2$ term, $M^{2} [30\pi \cos \phi -3\sin (3\phi )-60\phi \cos \phi +30\sin \phi ]/(16b^{3} )$, disagrees with Eq.~\eqref{uofphiSultana}. However, it remains unclear how the differentiation of Eq.~\eqref{dudphikds} leads to discrepancies in $M^2$ and $a^2$ terms. Moreover, both the $M^2$ and $a^2$ terms in the $(\mathrm{d}u/\mathrm{d}\phi)^2$ derived from Eq.~\eqref{uofphiSultana}, i.e., Eq.~\eqref{dudphi2sultana}, do not align with those in the original EOM.
\end{itemize} 

\section{Deflection angle with de Sitter background}
\label{desitterbackground}
Using de Sitter spacetime as the background (instead of Minkowski spacetime) leads to the vanishing of the $\Lambda$ term in the deflection angle for SdS spacetimes. This is presented in Refs.~\cite{arakida2012effect,arakida2018light,takizawa2022gravitational} in different forms:
\begin{itemize}   
    \item In Ref.~\cite{arakida2012effect}, the impact parameter, which refers to the distance between the lens and the undeflected light, is obtained by solving the equation $(\mathrm{d}u/\mathrm{d}\phi)^2=0$ with $M=0$. Thus, the so-called "undeflected" or "deflected" is relative to the spacetime where $M=0$ and $\Lambda\ne 0$, i.e., the de Sitter spacetime. In other words, as stated in Ref.~\cite{arakida2012effect}, $\Lambda$ is thoroughly absorbed into the definition of the impact parameter. 
    
    \item In Ref.~\cite{arakida2018light}, an "unperturbed reference line $C_\Gamma$" is introduced by projecting the geodesic in de Sitter spacetime vertically onto the subspace of the SdS spacetime. An angle called "the total deflection angle" with respect to the $C_\Gamma$ is then proposed. As mentioned in Ref.~\cite{arakida2018light}, the deflection angle is calculated by assuming the de Sitter spacetime as the background of SdS spacetime. Thus, it is not surprising that the conclusion---the light ray does not bend in de Sitter spacetime and the $\Lambda$ term disappears in SdS spacetime---is drawn in Ref.~\cite{arakida2018light}.

    \item In Ref.~\cite{takizawa2022gravitational}, despite adopting the definition Eq.~\eqref{defineangle}, the $\Lambda$ term still disappears. The authors denote the angle determined by Eq.~\eqref{defineangle} as $\alpha$ (distinct from the "$\alpha$" in Secs.~\ref{Staticityofsourceandobserver} and \ref{Calculation}). They define the deflection angle of light in SdS spacetime as $\alpha(M\ne 0, \Lambda\ne 0) - \alpha(M = 0, \Lambda\ne 0) $, which they refer to as "the deflection angle on the de Sitter background". Since the $\Lambda$ term is identical in both $\alpha(M\ne 0, \Lambda\ne 0)$ and $\alpha(M = 0, \Lambda\ne 0)$, it cancels out in the final result.
\end{itemize}

 \bibliographystyle{elsarticle-num} 
 \bibliography{refs-kds-revised}

\end{document}